\documentclass[prl,twocolumn,floats,aps,epsfig,nofootinbib,axodraw,amssymb,11pt]{revtex4}
\def\beq{\begin{equation}}
\def\eeq{\end{equation}}
\def\bea{\begin{eqnarray}}
\def\eea{\end{eqnarray}}

\def\journal#1#2#3#4{{ #1} {\bf #2}, #3 (#4)}

\def\prd{Phys. Rev. D}

\def\gev{\rm GeV}

\def\dl{\Delta L}
\def\mn{m_N^{}}
\def\Smm{S_{\mu\mu}^{}}

\def\pslash{\not{\hbox{\kern-4pt $p$}}}
\def\qslash{\not{\hbox{\kern-4pt $q$}}}
\def\lv{\not{\hbox{\kern-4pt $L$}}}
\def\lsim{\mathrel{\raise.3ex\hbox{$<$\kern-.75em\lower1ex\hbox{$\sim$}}}}
\def\gsim{\mathrel{\raise.3ex\hbox{$>$\kern-.75em\lower1ex\hbox{$\sim$}}}}
\def\ifmath#1{\relax\ifmmode #1\else $#1$\fi}

\usepackage{graphicx}
\usepackage{bm}
\begin{document}
\draft
\renewcommand{\thefootnote}{\arabic{footnote}}

%



\title{Signatures for  Majorana neutrinos at  hadron colliders}
\bigskip

\author{Tao Han$^{1,2,3}$\footnote{email: than@physics.wisc.edu} and 
Bin Zhang$^2$\footnote{email: zb@mail.tsinghua.edu.cn} }
\address{$^1$Department of Physics, University of Wisconsin, Madison, WI 53706,
U.S.A. \\
$^2$Center for High Energy Physics, Tsinghua University, Beijing 100084, P.R.
China\\
$^3$Institute of Theoretical Physics, Academia Sinica, Beijing 100080, 
P.R. China}

\begin{abstract}
The Majorana nature of neutrinos may only be experimentally verified
via {\it lepton-number} violating processes  involving charged leptons.
We explore the $\dl=2$ like-sign dilepton production
at hadron colliders to search for  signals of  Majorana neutrinos.
We find significant sensitivity for resonant production
of a Majorana neutrino in the mass range of $10-80$ GeV at the current
run of the Tevatron with 2 fb$^{-1}$ integrated luminosity,
and in the range of $10-400$ GeV  at the LHC with 100 fb$^{-1}$.

\end{abstract}

\maketitle


Neutrinos are arguably the most elusive particles  in the standard model
(SM) spectrum.  The evidence is strong that neutrinos
are massive and their flavors defined with respect to the charged
leptons oscillate \cite{review}, indicating the need of extension beyond
the SM. We  do not know the nature
of the mass generation and flavor mixing.  In particular, we are clueless
if neutrinos are of Dirac or Majorana type --- the former preserves the
lepton number ($L$), and the latter violates it by two units. Thus,
the unambiguous proof of the existence of  a Majorana neutrino is the
observation of  a {\it lepton-number} violation process, which would
have profound implications in particle physics, nuclear physics, and cosmology.
Since neutrinos are so weakly interacting and leave no trace in ordinary
detectors, the only appropriate signatures must involve charged leptons
via the charge-current interactions for a $\dl=2$ process.

The simplest extension of the Standard Model to include Majorana 
neutrinos is to introduce  $n$ right-handed SM singlet neutrinos 
$N_{a R}\ (a=1,2, \cdots, n)$, 
and $n\ge 2$ in order to generate at least two massive neutrinos.
Besides the Dirac masses $m_D^{}$ from the Yukawa interactions,  
there is also a possible heavy Majorana mass term
$\sum_{b,b'=1}^{n} \  \overline{{N^c}_{b L}}\ B_{b b'}\ N_{b' R} + h.c.$
The diagonalized mass terms  thus read
\vskip -0.5cm
\bea
{1\over 2}
\left(\sum_{m=1}^3 m^\nu_m \overline{\nu_{m L}} \nu^c_{m R} +
\sum_{m'=4}^{3+n} m^N_{m'} \overline{N^c_{m'L}} N_{m'R} \right) 
\nonumber
\eea
plus the Hermitian conjugate, 
with the mixing relations  between the flavors defined with respect to the
charged leptons $\ell$ and mass eigenstates
\bea
\nonumber
&& \nu_{\ell L} = \sum_{m=1}^3 U_{\ell m}\nu_{m L}+\sum_{m'=4}^{3+n} 
V_{\ell m'} N^c_{m'L},\\
&& U U^\dag + V V^\dag = I.
\nonumber
\eea
In the simplest incarnation without further flavor structure or new states,
the light neutrino masses $m^\nu_m$ are of the order of magnitude
$m_D^2/B$, while the heavy neutrino masses are $m^N_{m'} \simeq B$.
The corresponding mixing angles are of  $ V V^* \sim m^\nu_m / m^N_{m'} $,
and thus $UU^\dag\approx I$. However, we will take a phenomenological
approach toward the mass and mixing parameters without assuming any
relationship a priori.

In terms of the mass eigenstates, the gauge interaction Lagrangian 
can be written as 
\bea
\nonumber
{\cal L}
&=& -\frac{g}{\sqrt{2}} W^+_\mu \left(
\sum_{\ell=e}^\tau \sum_{m=1}^3
U^{*}_{\ell m}\  \overline{\nu_m} \gamma^\mu P_L \ell \right) + h.c. \\
\nonumber
&-&  \frac{g}{\sqrt{2}} W^+_\mu \left(
\sum_{\ell=e}^\tau \sum_{m'=4}^{3+n}
V^{*}_{ \ell m'}\ \overline{N^c_{m'}} \gamma^\mu P_L \ell  \right) + h.c. \\
&-& \frac{g}{2\cos_W}Z_\mu  \left( \sum_{\ell=e}^\tau \sum_{m'=4}^{3+n}
V^{*}_{\ell m'}\ \overline{N^c_{m'}}\gamma^\mu P_L \nu_{\ell} \right) + h.c.
\nonumber
\eea
where $P_L$ is the left-handed chirality projection operator.
There exist constraints on the heavy neutrino mass
and the mixing elements $V_{\ell m'}$.
Since we are interested in the collider searches, we consider
$m_N\gg 1\ {\rm GeV}$. By far, the strongest bound is from
the non-observation of the neutrinoless double-$\beta$ decay 
($0\nu \beta\beta$) \cite{nnbb}. 
It translates to a bound on the mass and mixing element 
 \begin{equation}
\sum_{N} \frac{\left| V_{e N}\right|^2}{m_N^{} }
<5\times 10^{-8} \ {\rm GeV}^{-1}.
\label{beta}
\end{equation}
The other relevant  constraints come from the LEP 
experiments \cite{L3,DELPHI-OPAL}, typically leading to
$\left| V_{\mu N}\right|^{2},\ \left| V_{\tau N}\right|^{2} \lsim 
10^{-4}-10^{-5}$ for $\mn \sim 5\ {\gev} - 80\ {\gev}$.

We first consider the heavy Majorana neutrino decay width.
The decay modes of a heavy Majorana neutrino are to a $W$ 
or a $Z$ or a Higgs boson plus a corresponding SM lepton  \cite{Higgs}. 
 The total width of a heavy Majorana neutrino goes like
\begin{eqnarray}
\nonumber
\Gamma_N \approx \left\{
\begin{array}{ll} \displaystyle
\sum_\ell \left|V_{\ell N}\right| ^{2} \frac{G_F m^3_N }{8}\ 
 & {\rm for}\ \ m_N > m_Z,m_H\\ [6mm]
  \displaystyle
 \sum_\ell \left|V_{\ell N}\right| ^{2} \frac{G_F^2  m^5_N}{10^3}\ 
& {\rm for}\ \ m_N \ll m_W .
\nonumber
\end{array}
\right.
\nonumber
\end{eqnarray}
It remains rather narrow even for $m_N\sim 1$ TeV for small mixing angles.
The branching ratios of heavy Majorana neutrino decay are
$Br(N\rightarrow \ell^- W^+)\simeq 
Br(N\rightarrow  Z\nu  )\simeq 
Br(N\rightarrow  H\nu )\simeq 25\% $.

\begin{figure}[tb]
\includegraphics[width=5.1truecm,clip=true]{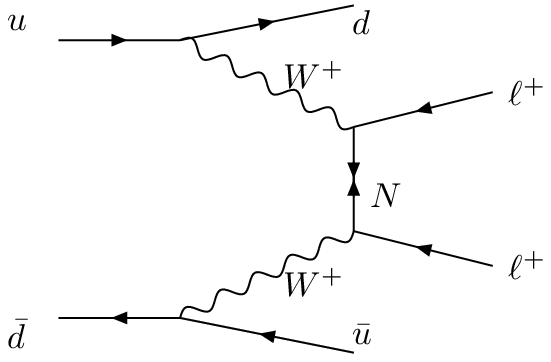}
\includegraphics[width=4.8truecm,   clip=true]{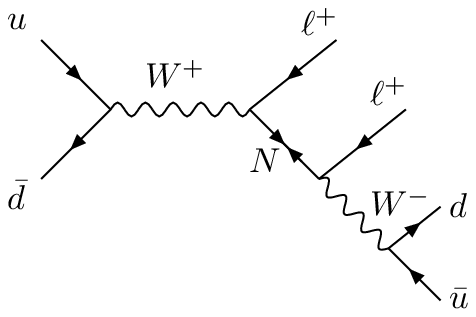}
\vspace{-0.5cm}
 \caption{ Feynman diagrams for $\dl=2$ processes induced by
a Majorana neutrino $N$ in $q\bar q'$ collisions.}
\label{Fig:feynmandiagram}
\end{figure}

Representative Feynman diagrams for $\dl=2$ processes induced by
 Majorana neutrinos in $q\bar q'$ collisions are depicted in
 Fig.~\ref{Fig:feynmandiagram}.
The contribution via $WW$ fusion shown by the first diagram is
the direct collider-analogue to the $0\nu\beta\beta$ and has been
discussed in  \cite{DDicus}. 
We reiterate that  the rate is very small due to the suppression  
of $\left|V_{\ell' N}\right|^{4}$.
We emphasize that a Majorana neutrino can be produced in resonance
as seen in the second diagram if kinematically accessible 
$\mn < \sqrt s$, thus substantially enhancing the production rate,
which is proportional to $\left|V_{\ell N}\right|^{2}$. This was noted
earlier in \cite{ACSV}, where only  $m_N > 100$ GeV was considered for
the $e^\pm e^\pm$ mode. We extend the calculation to include an
arbitrary mass $m_N^{}$ and general mixing for final state lepton flavors. 
The constraint of Eq.~(\ref{beta}) from $0\nu\beta\beta$
is very strong \cite{nnbb} and discourages the hope for signals 
involving an electron, which has been the focus so far in the
literature \cite{ACSV,DDicus,Ours}.
We thus propose the like-sign di-muons $\mu^\pm \mu^\pm$,
easier for detection than electrons  in hadronic collisions,
as the best signature for a heavy Majorana neutrino 
at both the Tevatron and LHC energies.
The final state $W$ boson is reconstructed in hadronic decay channels
because of the need of no neutrinos involved in the final state to assure
the unambiguous identification of $\dl=2$.

\begin{figure}[tb]
\includegraphics[width=8.5truecm,clip=true]{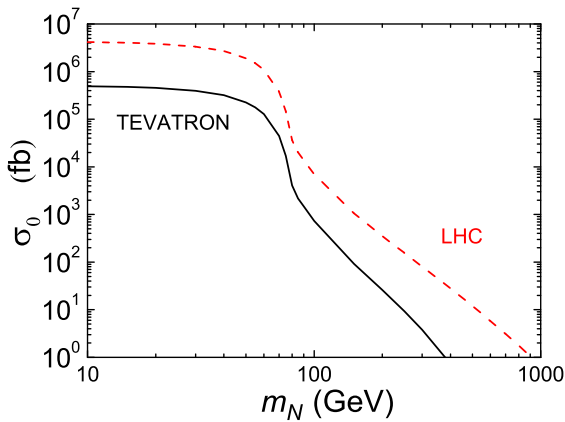}
\vspace{-1cm}
 \caption{ The ``bare cross section" $\sigma_0$ versus the heavy neutrino mass
 at the Tevatron and the LHC. }
\label{Fig:KvsMass}
\end{figure}

For the mass range of our current interest,  the width of the heavy neutrino
is small, so that the signal cross section can be approximated as
\bea
\nonumber
 \sigma(p\bar p&\rightarrow& \mu^\pm\mu^\pm W^\mp) \approx \\
 \sigma(p\bar p&\rightarrow& \mu^\pm N)Br(N\rightarrow \mu^\pm
W^\mp) \equiv S_{\mu\mu}\ \sigma_0,\quad \ 
\label{Effect}
\eea
where  $S_{\mu\mu}$ is the ``effective mixing parameter" of $N$ with a muon,  
defined by
\begin{equation}
 S_{\mu\mu}=\frac{\left|V_{\mu N}\right| ^{4}}
{\sum_\ell \left|V_{\ell N}\right| ^{2}}
\end{equation}
and $\sigma_0$ is ``bare cross section",
essentially independent of the mixing
parameters when the heavy neutrino decay width is narrow. 
We calculated the exact
matrix elements including all contributing diagrams and found
that the factorization of Eq.~(\ref{Effect}) is a good approximation.
Our results are shown in Fig.~\ref{Fig:KvsMass}.

\vskip 0.2cm
\noindent
\underline{The Tevatron}:
To be more realistic in estimating the signal observability,
we introduce the basic acceptance  on leptons and jets to simulate
the CDF/D0 detector coverage in the transverse momentum ($p_T^{}$)
and pseudo-rapidity ($\eta$)
\begin{eqnarray}
&& p_T(\mu)>5\  {\,\rm GeV}, \;\;\;\;\; \eta(\mu)<2.0,\\
&& p_T(j)>10\ {\,\rm GeV}, \;\;\;\;\; \eta(j)<3.0.
\end{eqnarray}
The important characteristics for the signal are the 
two well-isolated like-sign leptons, with no missing energy. 
We thus require the events to have $\mu$-jet separation and small missing energy
\begin{eqnarray}
\Delta R^{min}_{\mu j}> 0.5,\quad p\!\!\!\slash_T < 20\ \rm{GeV}.
\end{eqnarray}
 The fully reconstructed events should reflect the nature of an on-shell
$W$ of the final states either in $m_W^{}\approx m(jj)$ or $ m(jj\mu\mu).$
We thus select events
\begin{eqnarray}
 60\ {\rm GeV}<m_{\rm cluster} <100\ {\rm GeV},
\label{Masses}
 \end{eqnarray}
where $m_{\rm cluster}$ is the invariant mass of $W$
either in $jj$ or in  $jj\mu\mu$ final state.
Furthermore, the signal distribution should naturally present 
a peak  at $m(jj\mu )\approx \mn$. 

Although there is no SM process as a background with $\dl=2$,
there are processes that lead to like-sign leptons. Those include
$W^\pm W^\pm jj$ production via gluon exchange or EW gauge boson
exchange, with $W^\pm$ decaying leptonically; or $W^\pm W^\pm W^\mp$
with the unlike-sign $W^\mp$ decaying hadronically. It turns out that the leading
background is from a cascade decay of $t \bar t$ production:
$t \to \ell^+\nu\  b\ ,\ \bar t\to W \bar b \rightarrow jets\ \bar c\ \ell^+ $.
With the cuts imposed above, the background can be
essentially eliminated.

\begin{figure}[tb]
\includegraphics[width=8.2truecm,clip=true]{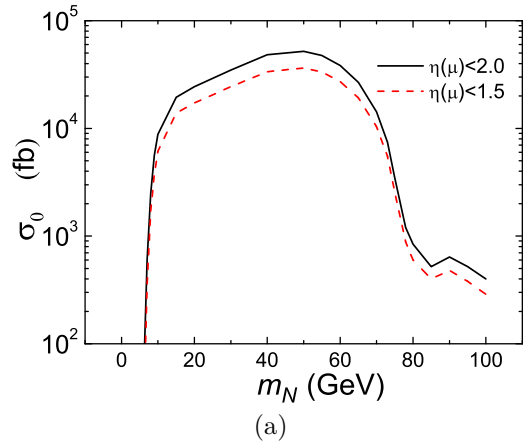}\vspace{-0.5cm}\hspace{-2cm}(a)
\includegraphics[width=8.2truecm,clip=true]{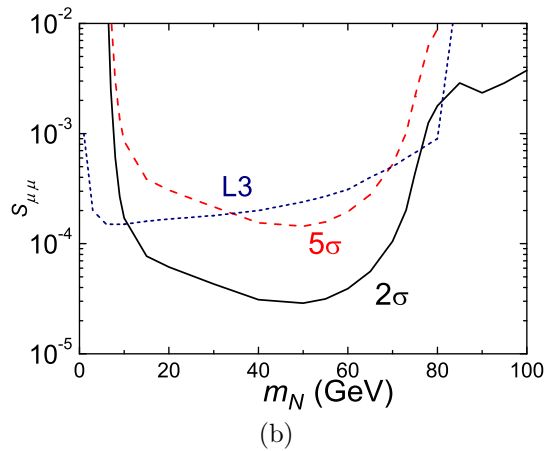}\vspace{-0.5cm}\hspace{-2cm}(b)
 \caption{Tevatron results: (a) $\sigma_0$ versus  $m_N$ after all the cuts with
 the two cases of
 muon rapidity acceptance from D0 and CDF; (b) Sensitivity on $S_{\mu\mu}$
 for  $2\ {\rm fb}^{-1}$.
  For comparison, the  $95\%$ bound from L3 search is  included.  }
\label{Fig:TeV}
\end{figure}

We smear the jet energy and the muon momentum
according to the Gaussian response of the detector.
We consider Poisson statistics for the low event rate
when calculating the signal significance. 
For instance, a $95\%\ (2\sigma$) bound on the signal
for no background would need a signal event rate
$N_S  = {\cal L}\ \sigma_0 (m_N)S_{\mu\mu}\geq 3$.
We show our final results at the Tevatron in Fig.~\ref{Fig:TeV},
the bare cross section $\sigma_0$ after all the cuts
and the sensitivity on the mixing parameter $\Smm$.
We see that there is significant sensitivity at the Tevatron in the
mass range of $\mn=10\ {\gev} - 80$ GeV. The mixing parameters 
can be probed to a few times $10^{-5}$ at a $2\sigma$ level with 2 fb$^{-1}$
integrated luminosity surpassing the L3 $95\% $ C.L.~bound \cite{L3}, 
and reaching about $10^{-4}$ at a $5\sigma$ level. While the Tevatron 
Run-II will be soon accumulating substantially more data, 
the search for a heavy Majorana neutrino 
will be particular interesting due to the fact that this would be
a low-background clean channel.

\vskip 0.2cm
\noindent
\underline{The LHC:}\
At the LHC, we adopt the judicious  cuts 
\begin{eqnarray}
&& p_T(\mu)>10\  {\,\rm GeV}, \;\;\;\;\; \eta(\mu)<2.5,\\
&& p_T(j)>15\ {\,\rm GeV}, \;\;\;\;\; \eta(j)<3.0, \\
&& \Delta R^{min}_{lj}> 0.5,\quad p\!\!\!\slash_T < 25\ \rm{GeV}.
\end{eqnarray}
A similar $W$ mass reconstruction is required as in Eq.~(\ref{Masses}).

\begin{figure}[tb]
\includegraphics[width=8.2truecm,clip=true]{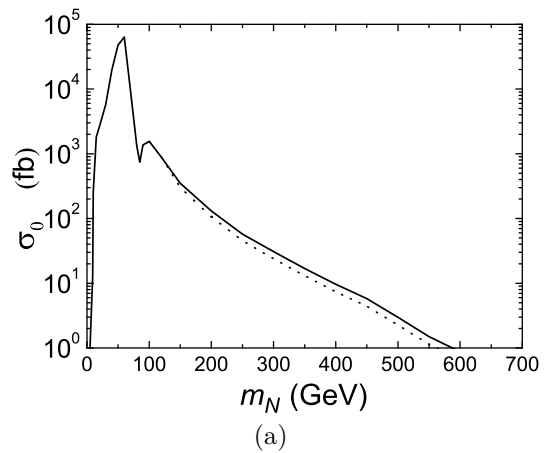}\vspace{-0.5cm}\hspace{-2cm}(a)
\includegraphics[width=8.4truecm,clip=true]{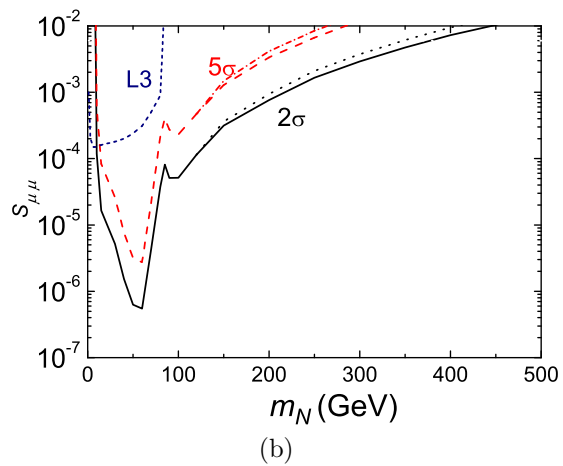}\vspace{-0.5cm}\hspace{-2cm}(b)
 \caption{LHC results: (a) $\sigma_0$ versus  $m_N$ after all the cuts;
 (b) Sensitivity on $S_{\mu\mu}$  for  $100\ {\rm fb}^{-1}$.
 For comparison, the $95\%$  bound from L3 search is  included. 
 The dotted lines tracing the curves illustrate the possible Higgs
 contribution of a mass 120 GeV.  }
\label{Fig:LHC}
\end{figure}

Along with the background processes studied for the Tevatron,
more channels are considered such as
$pp \rightarrow jjZZ$ and $pp \rightarrow jjZW$ that may fake
the like-sign dilepton signal when some particles are missing
from detection. Although most of backgrounds
can be made small  after the cuts, there are some events left
to contaminate the purity of the signal. The leading background is
$pp \rightarrow W^{\pm}W^{\pm}W^{\mp} $. The total background 
 after the cuts is about $(7-8)\times 10^{-2}$ fb. 
 One can further strengthen the signal significance by examining 
 the mass window 
\beq
0.8m_N<m(jj\mu)\approx \mn <1.2m_N.
\label{Statis}
\eeq
This would reduce the background below $0.04$ fb. 
Our  results after the cuts and the signal reconstruction of Eq.~(\ref{Statis}) 
are shown in Fig.~\ref{Fig:LHC},
for the bare cross section $\sigma_0$ and the sensitivity on
the effective  mixing parameter $\Smm$ with an integrated
luminosity 100 fb$^{-1}$.
We see that the mass range with significant sensitivity can be
extended to $\mn=10\ {\gev} - 400$ GeV. The mixing parameter 
can be probed to $10^{-3}$ at a $2\sigma$ level
 way beyond the L3 $95\% $ C.L.~bound \cite{L3}, 
and reaching about $10^{-6}$  in the low mass region at a $5\sigma$ level.
 The dotted lines closely tracing the curves illustrate the possible Higgs
 contribution to the total width $\Gamma_N$  with  a Higgs mass 120 GeV. 

It is only definitive to observe {\it lepton-number} violation processes 
to establish the Majorana nature of the neutrino masses, which would
have profound implications in particle and nuclear physics and cosmology. 
The experiments at the current Tevatron Run-II
and the LHC may have the opportunity to discover it via the distinctive
channels of like-sign dilepton production with no missing energy.
Hadron colliders may serve as the discovery machine for the mysterious
``sterile Majorana neutrinos".

\vskip 0.2cm
\noindent
{\it Acknowledgments:}
We thank A.~Atre and S.~Pascoli for collaboration on some related 
topics, and Y.-P. Kuang for participation at an earlier stage.
This work was supported in part by the U.S.~Department of Energy
under grant DE-FG02-95ER40896,
in part by the Wisconsin Alumni Research Foundation, and
in part by NSFC 10435040.


\begin{thebibliography}{000}

\bibitem{review} For recent reviews on neutrino masses and
flavor oscillations, see {\it e.g.},
V.~Barger, D.~Marfatia, and K.~Whisnant, 
Int.~J.~Mod.~Phys.~{\bf E12}, 569 (2003);
B.~Kayser, p.~145 of the Review of Particle Physics,
Phys.~Lett.~{\bf B592}, 1 (2004).

\bibitem{nnbb} For a recent review on neutrinoless double-$\beta$ decay,
see {\it e.g.}, S.~Elliott and J.~Engel,
J.~Phys.~{\bf G30}, R183 (2004); 
%
P.~Bamert, C.~P.~Burgess and R.~N.~Mohapatra,
  Nucl.\ Phys.\ B {\bf 438}, 3 (1995);
E. Nardi, E. Roulet, and D. Tommasini, Phys. Lett. B {\bf 344}, 225 (1995);
%
G. B\'elanger, F. Boudjema, D. London, and H. Nadeau, \journal{\prd} {53}{6292}{1996};
P.~Benes, A.~Faessler, F.~Simkovic, and S.~Kovalenko,
 Phys.~Rev.~{\bf D71}, 077901 (2005).

\bibitem{L3} 
L3 Collaboratioin: O. Adriani et al., Phys.~Lett.~{\bf B295}, 371 (1992);  
P.~Achard et al., Phys.~Lett.~{\bf B517}, 67 (2001).

\bibitem{DELPHI-OPAL} 
DELPHI Collaboration: P. Abreu et al.,
Z.~Phys.~{\bf C74}, 57 (1997), Erratum-ibid.{\bf C75}, 580 (1997); 
OPAL Collaboration: M.~Z.~Akrawy et al., Phys.~Lett.~{\bf B247}, 448 (1990).

\bibitem{Higgs} 
A.~Pilaftsis,  Z.~Phys.~{\bf C55}, 275 (1992).

\bibitem{DDicus} 
D.~Dicus, D.~Karatas, and P.~Roy, Phys.~Rev.~{\bf D44}, 
2033 (1991).

%
\bibitem{ACSV} 
A.~Datta, M.~Guchait, and A.~Pilaftsis, Phys.~Rev.~{\bf D50}, 3195 (1994);
F.M.L. Almeida, Y.A. Coutinho, J.A.M. Simoes and M.A.B. Vale,
Phys.~Rev.~{\bf D62}, 075004 (2000);
O. Panella, M. Cannoni, C. Carimalo, and Y.N. Srivastava,
Phys.~Rev.~{\bf D65}, 035005 (2002).

\bibitem{Ours} 
For a more comprehensive treatment  on $\dl=2$ processes
at hadron colliders and in rare decays, see {\it e.~g.},
A.~Atre, T.~Han, S. Pascoli, and B.~Zhang, to appear.

%


\end{thebibliography}
\end{document}